\documentclass[twocolumn]{aastex62}

\shortauthors{Madrid et al.}

\begin{document}


\title{{\sc A wide field map of intracluster globular clusters in Coma}}


\correspondingauthor{Juan P.\ Madrid }
\email{jmadrid@astro.swin.edu.au}

\author{Juan P.\ Madrid}
\affiliation{CSIRO, Astronomy and Space Science, PO BOX 76, Epping NSW 1710, Australia}
\affiliation{Gemini Observatory,  Southern Operations  Center, Colina El Pino s/n, La Serena, Chile}

\author{Conor R.\ O'Neill}
\affiliation{Gemini Observatory,  Southern Operations  Center, Colina El Pino s/n, La Serena, Chile}
\affiliation{Australian Astronomical Observatory, PO Box 915, North Ryde, NSW 1670, Australia}

\author{Alexander T.\ Gagliano}
\affiliation{Virginia Tech University, Blacksburg, VA 24061, USA}
\affiliation{Cerro Tololo Inter-American Observatory, Colina El Pino s/n, La Serena, Chile}
\affiliation{Los Alamos National Laboratory, P.O. Box 1663, Los Alamos, NM 87545, USA}

\author{Joshua R. Marvil}
\affiliation{CSIRO, Astronomy and Space Science, PO BOX 76, Epping NSW 1710, Australia}
\affiliation{National Radio Astronomy Observatory, 1003 Lopezville Road, Socorro, NM 87801, USA}


                         
\begin{abstract}

The large-scale distribution of globular clusters in the central region of the Coma cluster of galaxies
is derived through the analysis of Hubble Space Telescope/Advanced Camera for Surveys data.
Data from three different HST observing programs are combined in order to obtain a full surface 
density map of globular clusters in the core of Coma. A total of 22,426 Globular cluster candidates
were selected through a detailed morphological inspection and the analysis of their magnitude 
and colors in two wavebands, F475W (Sloan $g$) and F814W ($I$). The spatial distribution of globular 
clusters defines three main overdensities in Coma that can be associated with NGC\ 4889, NGC\ 4874, 
and IC\ 4051 but have spatial scales five to six times larger than individual galaxies. The highest 
surface density of globular clusters in Coma is spatially coincidental with NGC\ 4889. The most extended 
overdensity of globular clusters is associated with NGC\ 4874. Intracluster globular clusters also form clear 
bridges between Coma galaxies. Red globular clusters, which agglomerate around the center of the three main 
subgroups, reach higher surface densities than blue ones.

\end{abstract}



\section{Introduction}

The ability to detect extended stellar light with low surface brightness has changed
our understanding and perception of galaxies. \citet{mihos2005}
obtained an extremely deep optical image of the Virgo cluster, revealing a striking
and intricate web of intracluster light that has challenged our established view of 
the morphology of well studied galaxies such as M87, M86, and M84. Indeed, \citet{mihos2005}
revealed tails, bridges, and common envelopes between galaxies, a wealth of structure 
that had remained undetected until then. 

Similar efforts to image  faint  stellar structures ($\mu \sim 28$ mag)  have brought out tidal 
tails and stellar streams that show recent interactions of many galaxies 
\citep[e.g.][]{martinez2010,malin1997}. An interesting example of faint stellar structures 
in the Local Group revealed with deep wide-field imaging is given by 
the Pan-Andromeda Archaeological Survey \citep[PAndAS, e.g.,][]{mcconnachie2009, ibata2014, martin2014}. 
PAndAS unveiled, for instance, structures that are the remnants of dwarf galaxies likely 
destroyed by the tidal field of M31. Studies of the {\sc H\,i} line have also unveiled tidal 
tails and gas structures between interacting galaxies that are hidden to optical 
observations \citep[][among others]{yun1994,lopezsanchez2012}.

Similar to intracluster light, intracluster globular clusters (IGCs) are generally defined
as those globular clusters that are bound to the gravitational potential of a galaxy cluster
instead of a particular galaxy. Observational hints and theoretical predictions of the existence 
of IGCs could be found in the literature more than three decades 
ago  \citep[][among others]{forte1982,muzzio1984,white1987}. 

The unique sensitivity and resolution of the Advanced Camera for 
Surveys (ACS) on board the Hubble Space Telescope (HST) were instrumental in discovering 
and characterizing extragalactic globular cluster populations. First, with the WFPC2 and
then with the ACS, a number of IGC detections were reported in Abell 1185 \citep{jordan2003,west2011},
Virgo \citep{williams2007}, and Abell 1689 \citep{alamomartinez2013}. A tentative finding 
of IGCs in Fornax was also made by \citet{bassino2003} using 
ground-based data.

The first detection of a large population of IGCs was made with SDSS data by \citet{lee2010}, 
who mapped the large-scale structure of IGCs in Virgo. \citet{lee2010} 
found that globular clusters in Virgo define structures that are more extended than galaxies. 
From the maps made by \citet{lee2010} it became obvious that a large fraction of globular 
clusters in Virgo are actually bound to the galaxy cluster potential instead of individual 
galaxies. A clear bridge of globular clusters between M87 and M86 is also evident \citep{lee2010}.
\citet{durrell2014} expanded the work on Virgo using data from the Next Generation Virgo Cluster Survey. 

In this paper, we present the results of a search for globular clusters in the central 
region of the Coma cluster carried out using HST data taken with the ACS. The existence 
and characteristics of intracluster globular clusters in 
Coma were proven by \citet{peng2011} and we build on their work by obtaining a larger 
field of view that is nearly continuous for the core of Coma.

The Coma cluster of galaxies, a nearby Abell cluster, consists of a massive population of 
diverse galaxies \citep[e.g.][]{abell1977,colles1996}. It has been the focus of intense study 
due in part to its high concentration of gravitationally bound objects, its extremely bright 
spiral galaxies  and supermassive ellipticals NGC 4874 and NGC 4889 \citep{jorgensen1999,boselli2006}.
Its close proximity to the north galactic pole of the Milky Way also allows 
for deep imaging of Coma unobscured by intermediate gas, dust, or foreground stars. 

In the following sections, we discuss the HST observations we use, as well as our detection 
and selection criteria. We then present the color-magnitude diagram (CMD), the observed 
globular cluster luminosity function (GCLF), and the map of the large-scale structure formed by 
globular clusters in Coma. We obtain the surface brightness profiles of galaxies and compare 
them with the radial profiles of globular clusters. The spatial distribution of red and blue 
globular clusters is also shown. During this work, we create one of the largest globular 
clusters catalogs to date.

For this work a distance to the Coma cluster of 100 Mpc is adopted 
\citep[($m - M$) = 35.0 mag;][]{carter2008}. This value yields a scale of 23 pc 
per 0.05$\arcsec$, or one ACS pixel.

\begin{figure}
 \epsscale{1.15}
  \plotone{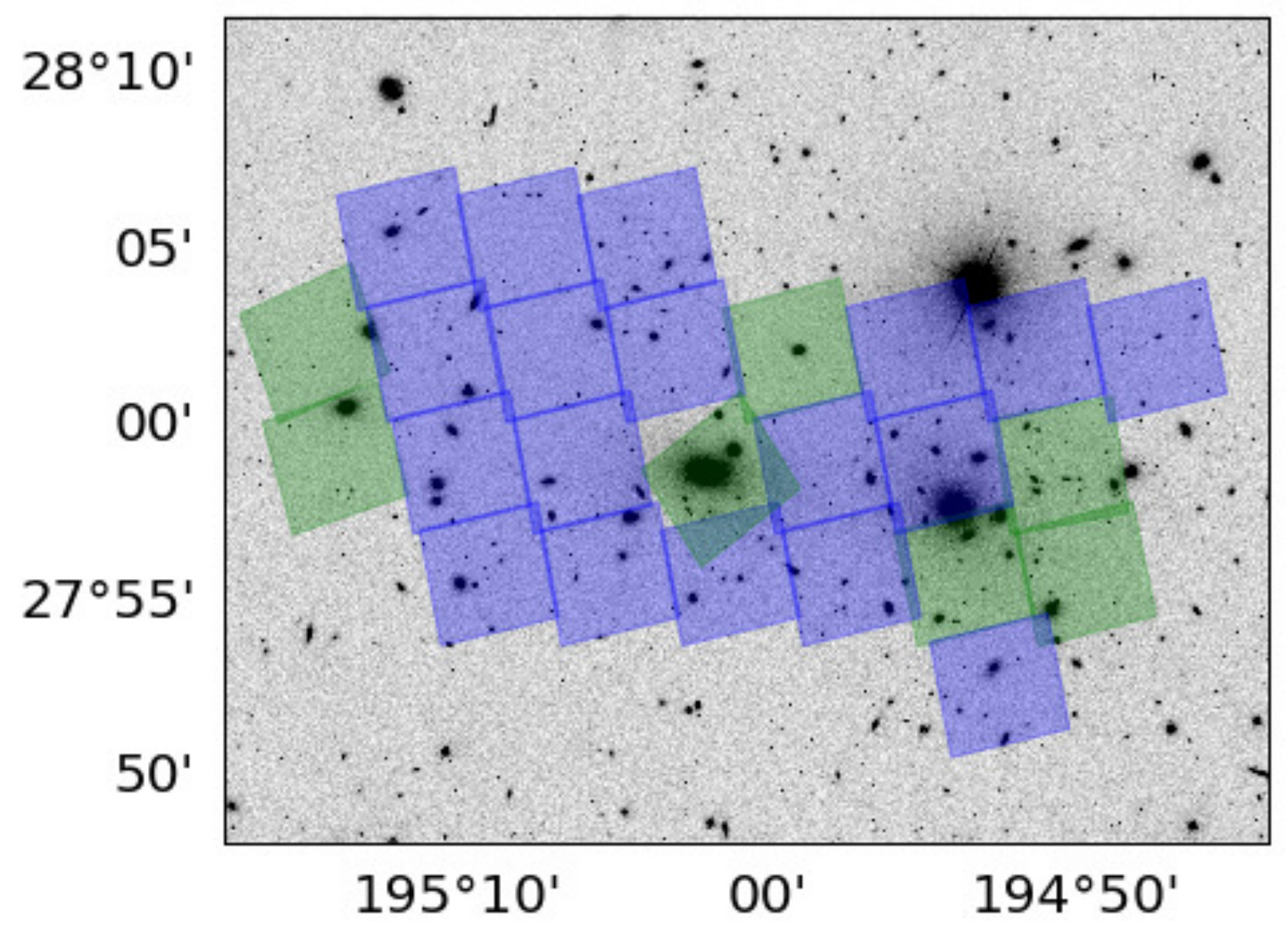}
 \caption{ACS pointings overlaid on a SDSS $g$-band image. The blue pointings were part of the study
 of Peng et al.\ (2011). For this work we include both blue and green pointings achieving a better
 coverage of the core of Coma. The green pointings were obtained under HST programs GO 11711 and GO 12918.\\
 \label{acspointings}}
\end{figure}
 


\begin{deluxetable}{cccc}
\tablecaption{Properties of Coma galaxies and GC surface density\label{tbl-1}} 
\tablehead{
\colhead{Galaxy} & \colhead{$m_V$} & \colhead{Velocity} & \colhead{GC Surface}\\
\colhead{Name} & \colhead{} & \colhead{} & \colhead{Density}\\
\colhead{} & \colhead{(mag)} & \colhead{(km/s)} & \colhead{(N/$\arcsec^2$)}\\
}
\startdata
IC\ 3973  & 14.37 & 4710 & 0.022 \\
IC\ 3976  & 14.70 & 6789 & 0.016 \\
IC\ 3998  & 14.57 & 9419 & 0.049 \\ 
IC\ 4011  & 15.12 & 7269 & 0.048 \\ 
IC\ 4012  & 14.94 & 7237 & 0.010 \\
IC\ 4021  & 14.85 & 5732 & 0.005 \\
IC\ 4026  & 14.59 & 8176 & 0.019 \\
IC\ 4030  & 15.40 & 7004 & 0.008 \\
IC\ 4033  & 15.21 & 7715 & 0.012 \\
IC\ 4040  & 14.84 & 7840 & 0.018 \\
IC\ 4041  & 14.35 & 7087 & 0.023 \\
IC\ 4045  & 13.94 & 6968 & 0.031 \\ 
IC\ 4051  & 13.63 & 8793 & 0.294 \\
NGC\ 4867 & 14.46 & 4822 & 0.034 \\
NGC\ 4869 & 13.76 & 6859 & 0.049 \\
NGC\ 4871 & 14.14 & 6795 & 0.080 \\
NGC\ 4872 & 14.41 & 7180 & 0.101 \\
NGC\ 4873 & 14.11 & 5824 & 0.137 \\
NGC\ 4874 & 11.68 & 7168 & 0.239 \\
NGC\ 4875 & 14.65 & 8008 & 0.048 \\
NGC\ 4876 & 14.39 & 6701 & 0.011 \\
NGC\ 4882 & 13.86 & 6371 & 0.283 \\
NGC\ 4883 & 14.35 & 8151 & 0.017 \\
NGC\ 4889 & 11.49 & 6446 & 0.477 \\
NGC\ 4894 & 15.19 & 4638 & 0.024 \\  
NGC\ 4898 & 13.48 & 6661 & 0.049 \\
NGC\ 4906 & 14.10 & 7521 & 0.059 \\
NGC\ 4908 & 13.18 & 4903 & 0.063 \\
\enddata
\tablecomments{Properties of Coma galaxies and globular cluster surface density.
Column (1): notable galaxies in Coma; 
column (2): galaxy total $m_V$ in magnitudes from de Vaucouleurs et al.\ (1991); 
column (3): radial velocity in km/s from NED;
column (4): surface density of  globular clusters at the position of the galaxy in numbers 
per arcsecond$^2$.}
\label{galaxytable}
\end{deluxetable}


\section{Advance Camera for Surveys Observations}

We use Hubble Space Telescope  Advance Camera for Surveys Wide Field Channel 
(ACS/WFC) data. Part of these public data were obtained during the ACS Coma Cluster Treasury Survey
\citep[GO 10861;][]{carter2008}. We use the data products associated with data release 2.2 obtained 
from the Mikulski Archive for Space Telescopes at the STScI. We analyze the two bands obtained during the
survey: F475W (Sloan $g$) and F814W (Cousins $I$). The exposure times for these data are 2677 s and 1400 s 
for the F475W and F814W filters, respectively. Further details of the ACS Survey on Coma 
can be found in \citet{carter2008} and \citet{peng2011}. The photometric properties of these
data, such as limiting magnitude, were presented by \citet{hammer2010}.

The ACS Coma cluster survey was planned to spatially cover the entire central region of Coma, but 
due to an electronics failure of the camera, the survey was not fully completed, leaving areas 
without imaging data. To compensate for these data gaps, additional archival HST pointings 
were analyzed. These are the green pointings in Fig.\ 1. These additional pointings were obtained with the 
same filters under observing programs GO 11711 \citep[PI: J.\ Blakeslee;][]{cho2016} and GO 12918 
\citep[PI: K.\ Chiboucas;][]{harris2017}. 
By combining the imaging data of the programs above, we are able to 
finally study the central region of Coma as originally planned by the Coma Cluster Survey. 
A total of 25 ACS pointings were studied; each pointing has two filters. Compared with the study of 
\citet{peng2011}, we added seven additional pointings. These new pointings are crucial, as they
include NGC 4889 and IC 4051. NGC 4889 is one of the two giant ellipticals in the core of Coma
and its brightest galaxy (see Figure \ref{acspointings} and Table 1). IC 4051 is another giant elliptical 
on the outskirts of Coma that is known to have a large population of globular clusters \citep{woodworth2000}.

The definitive advantage of using high-resolution HST data is the ability to use morphology as
an additional parameter to discriminate between globular clusters and other objects. Indeed, 
morphological information has been successfully used to differentiate between globular clusters
and background galaxies at the distance of Coma \citep{peng2011}. A subsection of this data 
set was also used to successfully identify ultra-compact dwarfs (UCDs) based on their photometry 
and morphology \citep{madrid2010}.


\section{Detection and selection of globular cluster candidates}

Source detection is carried out using the task \texttt{find} within \texttt{daophot}. The final 
list of globular clusters candidates is achieved by analyzing their color, magnitudes, and size.
A detailed account of the selection processes involved in creating the final list of globular 
cluster candidates is given in the appendix.

Through detailed visual analysis of the properties of the candidates we produced 
a final list of globular cluster that is  virtually free of contaminants such as background galaxies and 
artifacts. We stress that all globular clusters in the final list of candidates were validated 
through visual inspection by displaying the detections on the screen and scanning them on each 
image, and in both filters. We thus build a master catalog of 22,426 direct detections of globular 
cluster candidates in Coma.

\section{Photometry and Color-Magnitude Diagram}

Flux measurements are derived using \texttt{phot} within \texttt{daophot}.
Aperture photometry is carried out with \texttt{phot} using an aperture 
radius of 4 pixels. Aperture correction is applied using the prescription of \citet{sirianni2005}.
We adopt a value of $E(B −- V)$ = 0.009 mag for the foreground galactic extinction. Photometric 
zero-points are obtained from the updated ACS zero-point tables maintained on the STScI website; we use 
ZP$_{F475W}$=26.131 mag and ZP$_{F814W}$=25.504 mag.

\begin{figure}
 \epsscale{1.2}
 \plotone{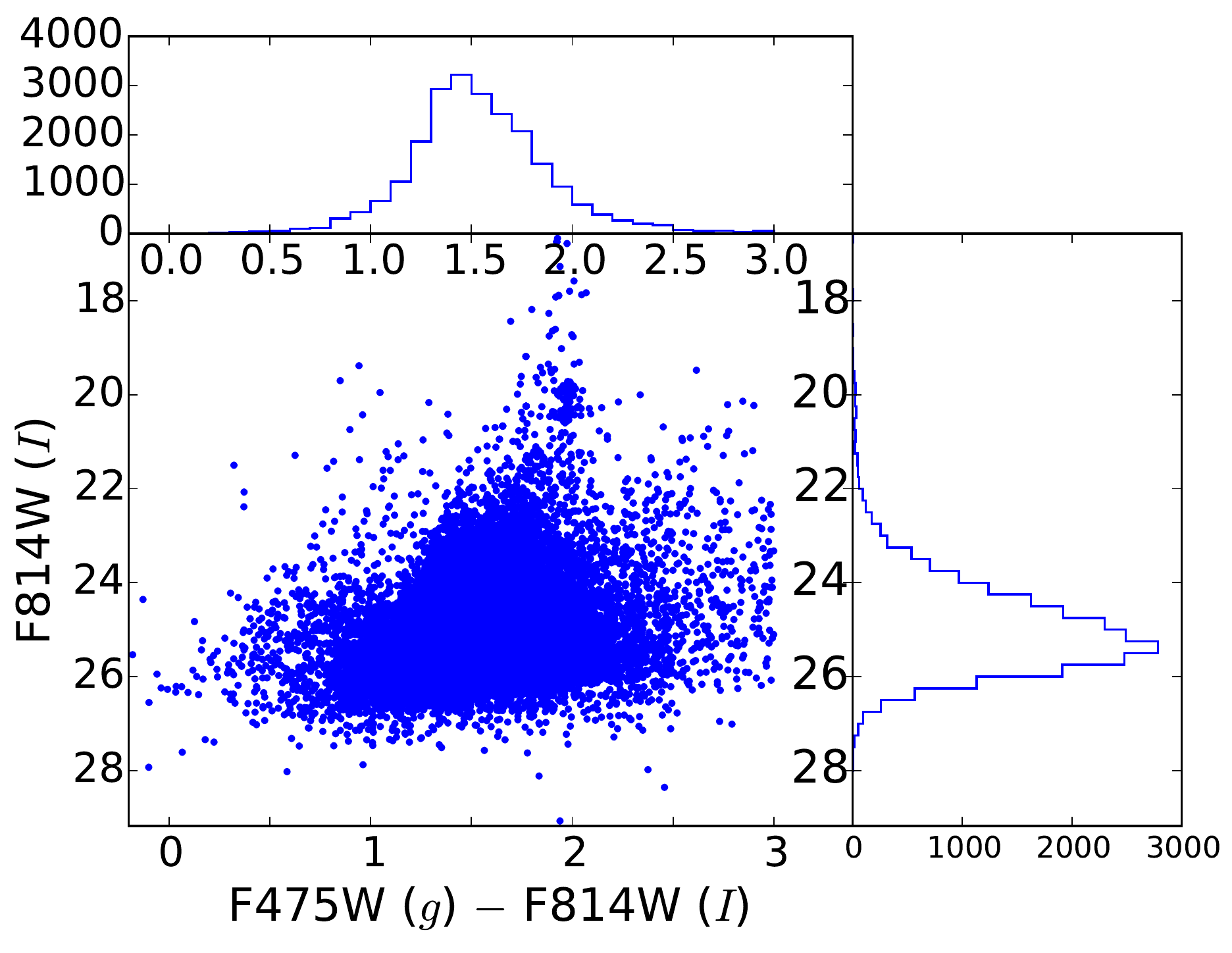}
 \caption{Color-magnitude diagram made with the final catalog of globular cluster candidates.
 The upper panel shows the color histogram, while the right panel shows the magnitude histogram or 
 luminosity function.\\
 \label{cmd}}
\end{figure}

The CMD made with the final list of 22,426 globular cluster candidates is 
shown in Figure \ref{cmd}.  This CMD is in good agreement with the magnitude and color distributions found 
in earlier work done with the HST on Coma globular clusters \citep{harris2009,madrid2010,chiboucas2011}.
A color histogram is presented above the CMD: 98.3\% of all the globular cluster
candidates have colors in the range $0.5<(F475W-F814W)<2.5$.

\section{Globular Cluster Luminosity Function}

\begin{figure}
 \epsscale{1.2}
 \plotone{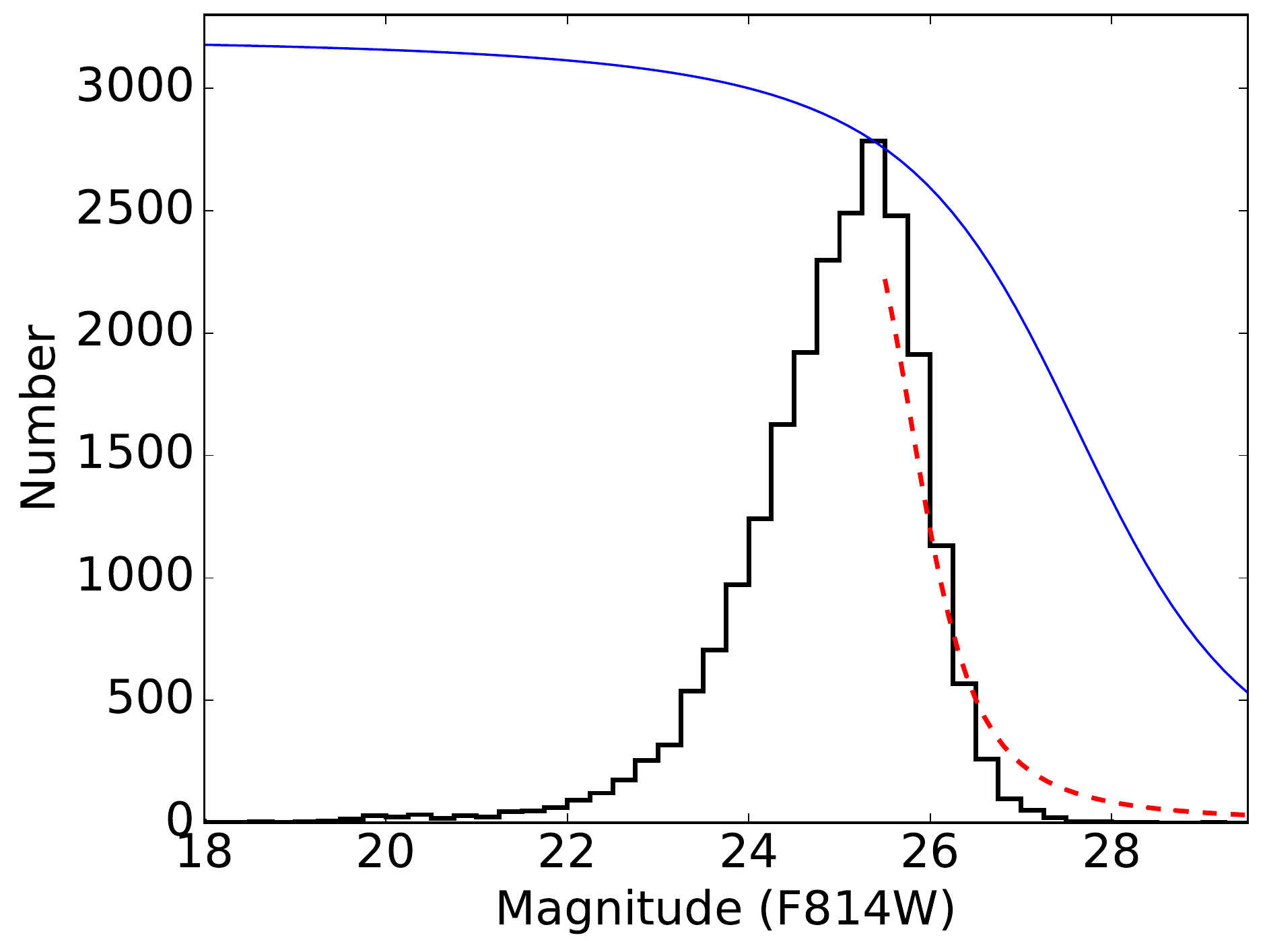}
 \caption{Observed globular cluster luminosity function (solid black histogram), and 
 ``completeness function"  of our 
 source finding routine (blue solid line). The red dashed line shows the fraction of detections that successfully 
 pass our selection criteria. For instance, at m$_{F814W}=25.0$ mag, 55\% of all detected sources
 are classified as globular cluster candidates. The rejection rate increases with fainter magnitudes;
 at  m$_{F814W}=26.0$ mag, only 13\% of detections pass the selection criteria.\\
 \label{gclf}}
\end{figure}

As mentioned by Peng et al.\ (2011), these data are too shallow to derive direct parameters for 
the GCLF. The histogram on the right hand side of the CMD is the 
observed GCLF in the F814W filter, as defined by our master catalog of candidates, reproduced in Figure \ref{gclf}. 
The peak of the histogram  in Figure \ref{gclf} is $\mu_{F814W}$ = 25.5 mag ($M_{F814W}$=-9.5~mag). Most globular 
clusters (99.7\%) have magnitudes in the range  $20 \le mag \le 28$ which translates into the absolute magnitude 
range of $-15 \le M_{F814W} \le -7$.

The turnover magnitude of the GCLF around bright ellipticals in Coma was determined to be about a magnitude
fainter than the peak of the histogram in Figure \ref{gclf}. Using HST/WFPC2 with longer exposure times (of up to 
31,200 seconds per filter), \citet{harris2009} found a turnover magnitude  of $M_{V,TO}\sim -7.5$ mag for globular 
clusters in Coma. In the analysis of the GCLF  in Virgo and Fornax 
\citet{villegas2010} achieved results similar to those of \citet{harris2009} in Coma.

The effectiveness of our detection algorithm is estimated by injecting artificial stars into the ACS files using the 
\texttt{pyraf} task \texttt{addstar} within \texttt{daophot}. Artificial stars with magnitudes between 
$m(F814W)=9$ mag and $m(F814W)$=35 mag are added. One hundred artificial stars per magnitude bin of 0.5 mag 
are injected into the images. Adopting these values avoids overcrowding. We thus measure 
the fraction of artificial stars that are successfully recovered. We adopt the functional form of the 
completeness function defined by \citet{harris2009}:

\begin{equation}
\label{eqn:completeness}
f = 0.5\Bigg(1-\frac{\alpha(m-m_0)}{\sqrt{1+\alpha^2(m-m_0)^2}}\Bigg)
\end{equation}

where $m_0$ is the completeness limit at which $f$ reaches 50\%, and $\alpha$ is the slope, 
which defines the rate at which $f$ declines as it passes through $m_0$ \citep{harris2009}. 
This function is plotted in Figure \ref{gclf} as a solid blue line with the following parameters:
$m_0=27.62$, and $\alpha=0.48$.


\begin{figure*}
\epsscale{1.2}
 \plotone{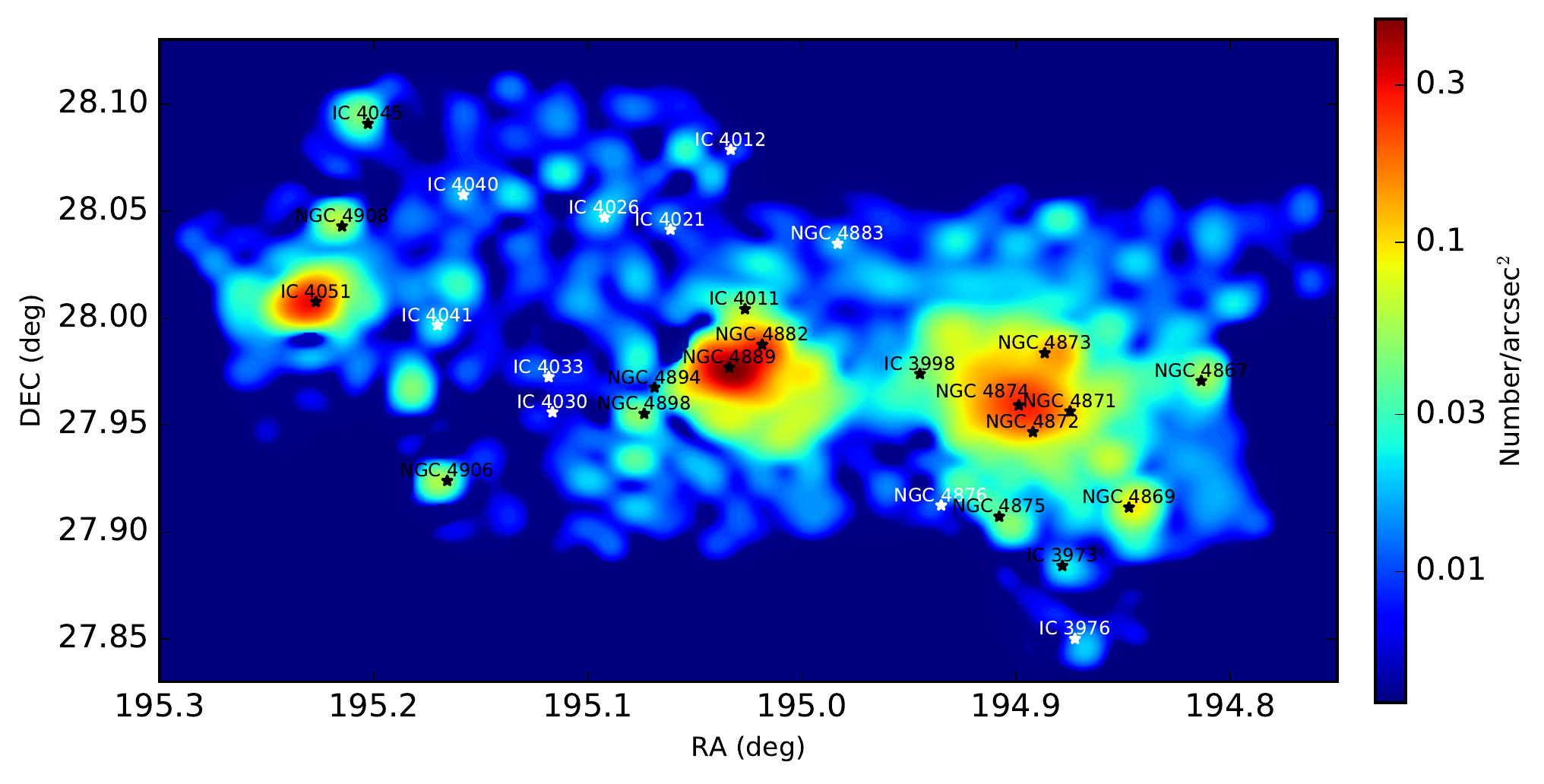}
 \caption{Surface density of globular clusters in the central regions of the Coma cluster. This heat map 
 defines three main overdensities of globular clusters in Coma. The highest density of globular clusters in 
 Coma is cospatial with NGC\ 4889. The most extended agglomeration of globular clusters can be associated
 with NGC\ 4874. A third overdensity is associated with IC\ 4051. North is up and east is left.
 \label{heatmap}}
\end{figure*}



The ``completeness function" derived above, only represents our ability to detect sources
with our source finding algorithm. As detailed in the Appendix, only $\sim$25\% of point 
sources initially detected on the HST images are classified as globular cluster candidates.

In order to illustrate the impact of our selection criteria on the observed globular cluster luminosity 
function we plot in Figure \ref{gclf}, as a red dashed line, the fraction of detections that successfully 
pass our selection criteria and are classified as globular cluster candidates. This red dashed line has 
the same functional form as Eq.~\ref{eqn:completeness}, with the following parameters: $m_0=25.80$, and 
$\alpha=1.35$. At m$_{F814W}=25.0$ mag, 55\% of all detected sources are classified as globular cluster 
candidates. The rejection rate increases with fainter magnitudes, at m$_{F814W}=26.0$ mag, only 13\% of all 
detections pass the selection criteria. The design of these observations and our conservative selection criteria 
make the observed luminosity function presented in this section unsuitable for a detailed analysis of the 
GCLF parameters.


UCDs are considered to have magnitudes of $M_V \le -11$ mag \citep{mieske2006}.
If we assume $(V-I)=1.1$~mag and the distance modulus of 35.0 mag quoted above, all sources in the master catalog
with $m_{F814W} < 22.9$ mag could be considered UCD candidates. Our catalog contains 867 UCD candidates, 
or 3.9\% of the sample.


\begin{figure}
        \centering
        \begin{tabular}{c} 
                \includegraphics[scale=0.42]{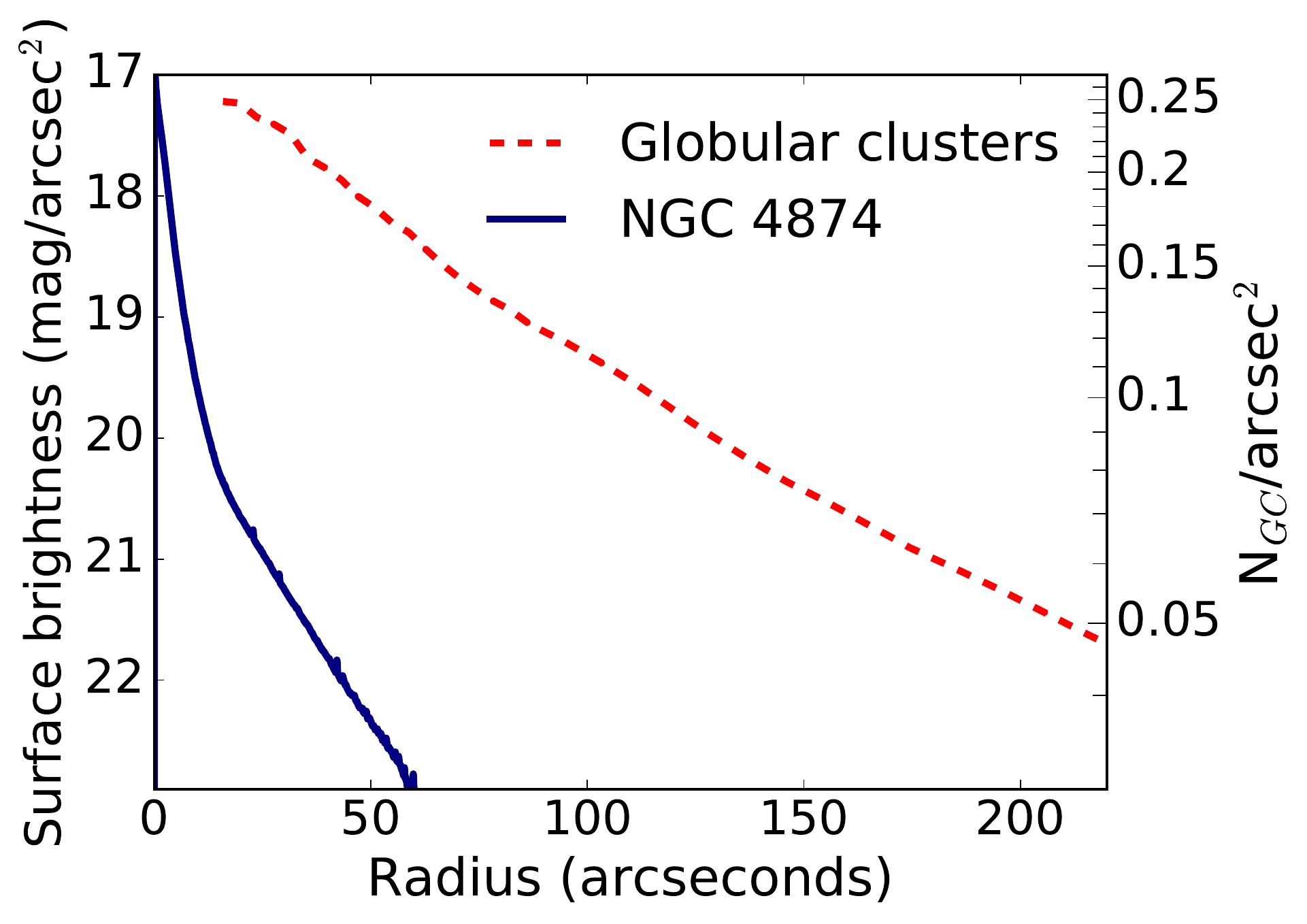}\\ 
                \includegraphics[scale=0.42]{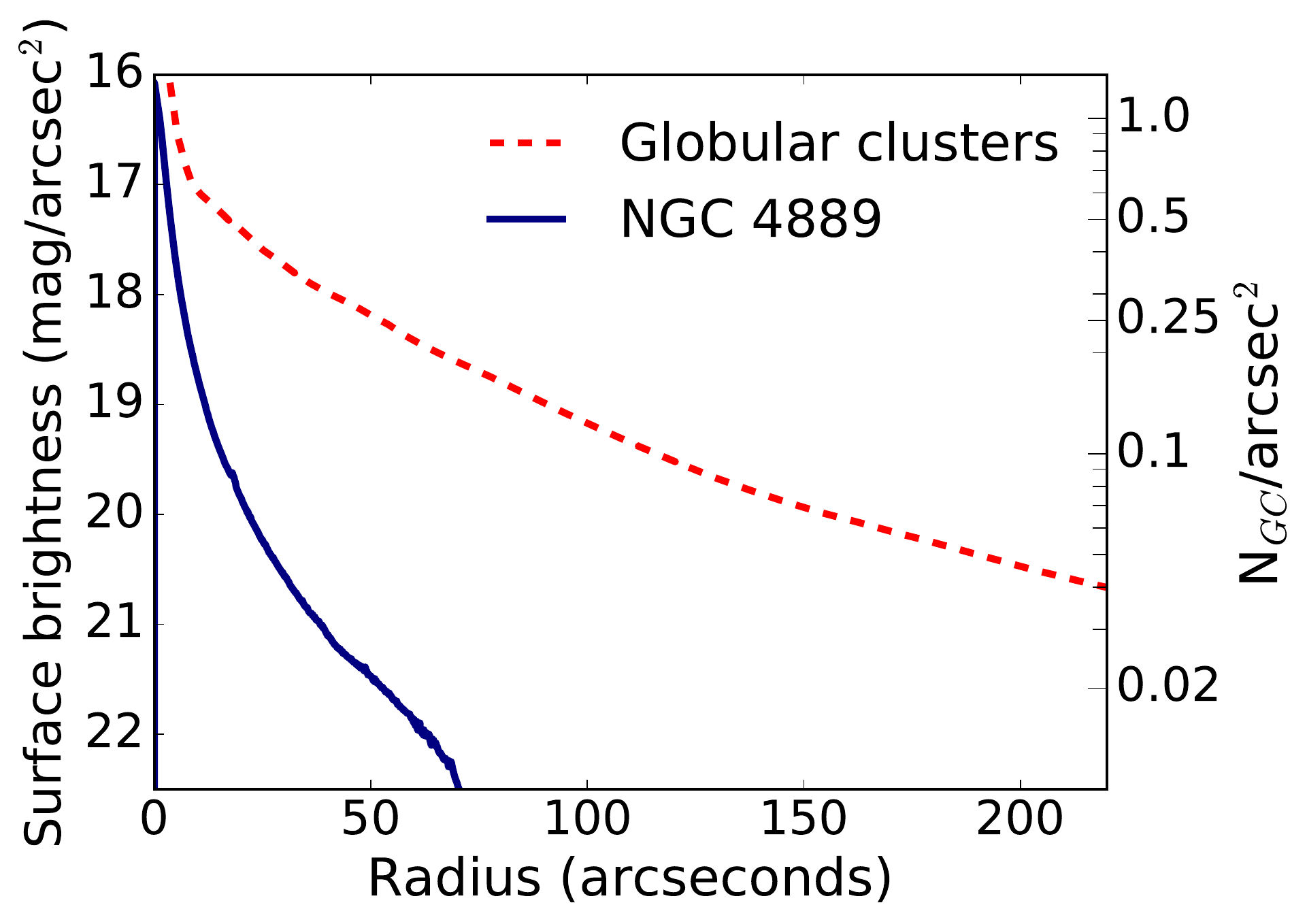}\\           
                \includegraphics[scale=0.42]{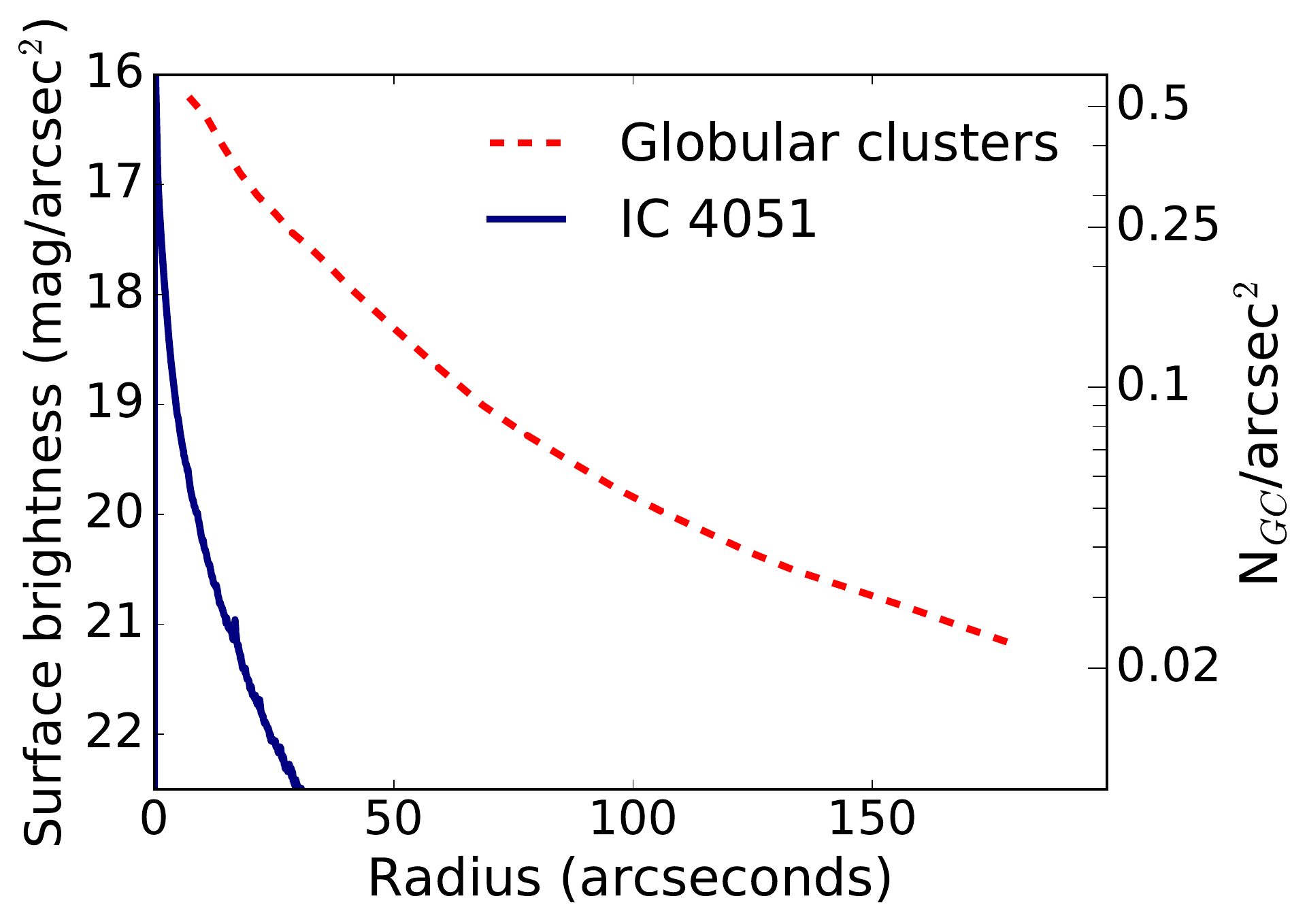}\\
        \end{tabular}
    \caption{Surface brightness profiles of NGC 4874, NGC 4889, IC 4051,
    and radial density profiles of globular clusters associated with these three
    overdensities. Globular clusters have radial distributions five to six times 
    more extended than the light of individual galaxies -- see also Table 2.\\
}
\label{figprofiles}
\end{figure}

\section{Wide-Field Map of Globular Clusters in Coma}

Surface density (or heat) maps are created using our final master catalog of globular clusters. 
These are made using the \verb|contourf| routine of the \verb|Matplotlib| library in \verb|Python|.
The contours use the \verb|ndimage.zoom| routine from the \verb|Scipy| library to interpolate 
and smooth the values. We use a spline interpolation of 4th order and a rescaling factor of two along 
each axis.

The heat map shown in Figure \ref{heatmap} shows that instead of simply belonging to a single parent 
galaxy, globular clusters outline the existence of a large-scale structure consistent with three 
main agglomerations. 

NGC\ 4874, NGC\ 4871, NGC\ 4872, and NGC\ 4873 appear to share a common envelope in one of the 
main agglomerations of globular clusters. Similarly, NGC\ 4889 and NGC\ 4882 are enshrouded by one 
common subgroup of globular clusters. In the core of Coma, that globular clusters could be associated 
with several large galaxies instead of a single one. A third, less populous, agglomeration of globular 
clusters can be associated with IC\ 4051, which defines the third peak of the surface density of 
globular clusters.

The globular clusters associated with the IC\ 4051 overdensity form a rather symmetrical and
elongated elliptical shape. The morphology of the two other overdensities located around the 
two cD galaxies is far more complex. These structures are clearly irregular and asymmetric. 

Moreover, the distribution of globular clusters is different from the distribution of galaxies. 
There is a clear bridge of globular clusters between NGC\ 4889 and NGC\ 4874. This bridge was 
already visible in the work of Peng et al.\ (2011). The globular cluster overdensity cospatial 
with NGC\ 4889 extends southwest of NGC\ 4889 into a region where no large galaxies are present. 

Table \ref{galaxytable} lists the most notable galaxies in Coma, along with their apparent magnitudes and 
radial velocities in km/s. This table also lists the value of the heat map at the location 
of each galaxy. The largest surface density of globular clusters in Coma,  0.477 GC/arcsec$^2$, 
is spatially coincident with one of the two cD galaxies: NGC 4889. Interestingly, the second 
highest surface density corresponds to IC\ 4051 instead of the other cD, NGC\ 4874. 

By studying a much wider field of view than \citet{peng2011} we are able to reveal sections 
of the core of Coma where the surface density of globular clusters falls to $\sim$0. Figure \ref{heatmap} 
shows that NGC\ 4906 is isolated from the large-scale structure of globular clusters, despite 
continuous imaging coverage. Another section of the map virtually free of globular clusters is located 
between IC\ 4033 and IC\ 4041.

The heat map on Figure \ref{heatmap} shows that several bright galaxies are located in regions with low 
surface-density of globular clusters, i.\ e.\  have globular cluster systems with low numbers of members.
Five clear examples of those bright galaxies associated with low numbers of globular clusters are
IC 4021 ($m_V=14.85$ mag), IC\ 4030 ($m_V=15.40$ mag), IC\ 4033 ($m_V=15.21$ mag), IC\ 4040 ($m_V=14.84 $ mag),
and NGC\ 4876 ($m_V=14.39$ mag). 


\begin{deluxetable}{lcc}
\tablecaption{Effective Radii\label{tbl-2}} 
\tablehead{
\colhead{Galaxy} & \colhead{Galaxy Surface} & \colhead{Globular Clusters}\\
\colhead{} & \colhead{Brightness Profile} & \colhead{Radial Profile}\\
\colhead{} & \colhead{(arcsec)} & \colhead{(arcsec)}\\
}
\startdata
NGC\ 4874 & 19.4 & 105.0 \\
NGC\ 4889 & 15.3 & 91.0 \\
IC\ 4051  & 8.2 &  45.6 \\
\enddata
\tablecomments{
Column (1): galaxy name; 
Column (2): effective radii of the galaxy's surface brightness profile (arcsec);
Column (3): effective radii of globular clusters radial profiles (arcsec).}
\label{tableprofiles}
\end{deluxetable}



\begin{figure}
\epsscale{1.2}
 \plotone{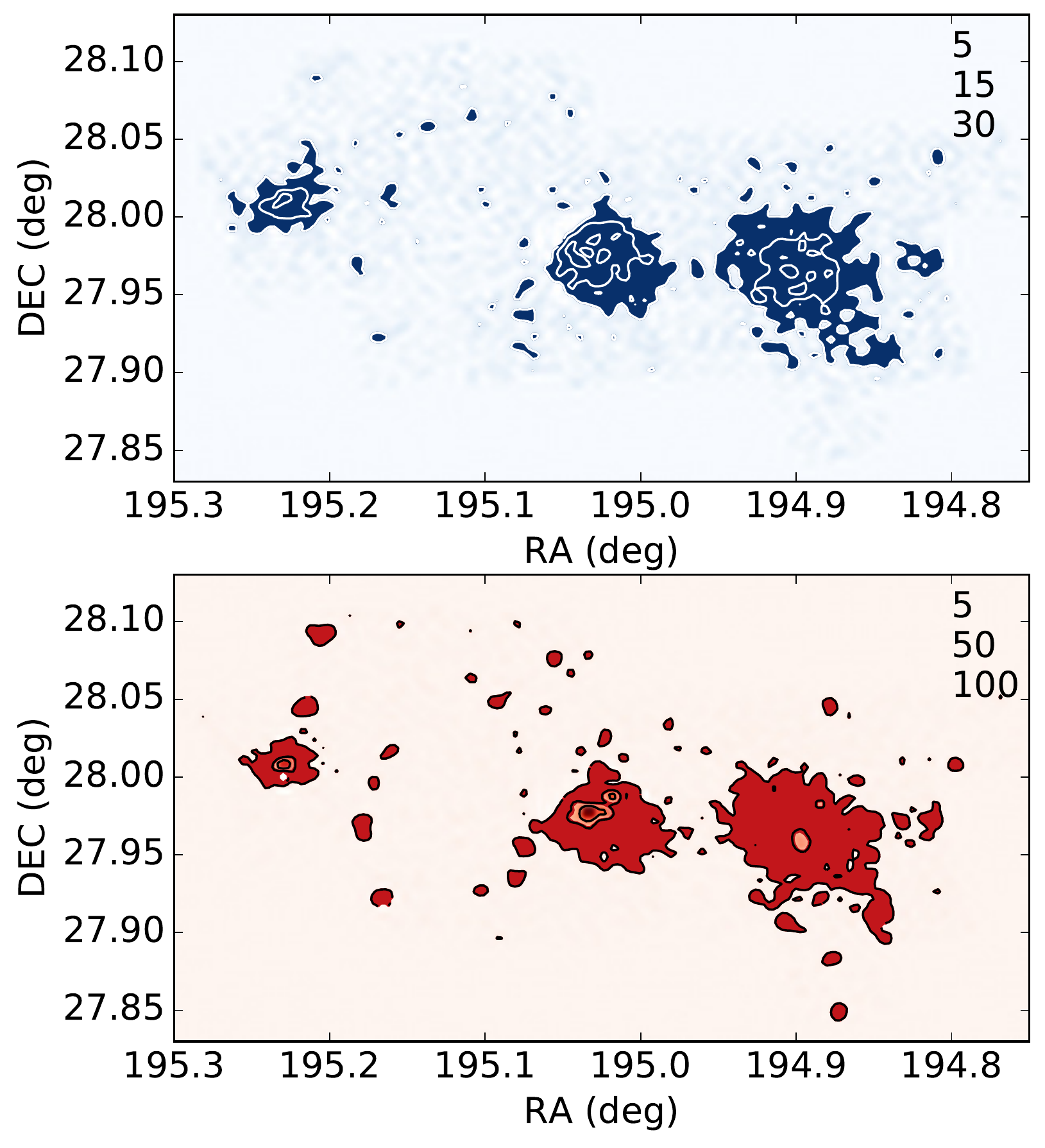}
 \caption{Surface density of blue ({\it top}) and red ({\it bottom}) globular clusters in the central 
 region of the Coma cluster. The red globular clusters have higher surface densities than blue ones. Isocontour 
 values are given in the top right hand corner. For a comparison with simulations see \citet{ramos2018}.
 \label{redblue}}
\end{figure}

\subsection{Spatial scales of globular cluster overdensities}

We derive the radial profiles for the surface brightness of NGC\ 4874, NGC\ 4889, IC\ 4051,
and their associated globular clusters, as shown in Figure \ref{figprofiles}. In order to quantify the 
spatial scales of both galaxy light and globular clusters, we fit a single S\'ersic model to 
their radial profiles. We should note that an exhaustive analysis of the surface brightness 
profile for these galaxies is beyond the scope of this paper. 
We fit a  S\'ersic model with the aim of obtaining a characteristic effective radius only. 
The effective radius of the S\'ersic model is derived using the prescriptions of \citet{graham2005}.
The results of this analysis are given in Table \ref{tableprofiles}. 

The spatial scales defined by the distribution of globular clusters are larger than the radius of 
any individual galaxy. As shown in Figure 5 and Table 2, globular clusters associated with 
NGC 4874 have an extended spatial structure with an effective radius that is more than 
five times the effective radius of the galaxy light. The NGC 4874 overdensity of globular
clusters is the largest in the core of Coma -- larger than the overdensity associated with
NGC 4889 or IC 4051.\\

\section{Distribution of {\it Red} and {\it Blue} globular clusters}

Globular clusters are often classified as a function of their color as
{\it red} or {\it blue}. These colors are in turn associated with 
metallicity, with blue globular clusters considered metal-poor and 
red deemed metal-rich \citep{zinn1984,larsen2001}. 
Several studies of individual galaxies have shown that blue globular
clusters have a more extended spatial distribution than red ones \citep{kundu1999,larsen2001}.
\citet{bassino2006} confirmed that red and blue globular clusters have different radial 
distributions at large distances ($\sim$200 kpc) from their host galaxies by studying 
Fornax (NGC 1399) using wide-field photometry.

\citet{lee2010} showed that blue IGCs form a more extended 
distribution than red ones in the Virgo Cluster, as did \citet{bassino2006} with Fornax. 
We investigate the distribution of red and blue globular clusters in Coma by separating 
globular clusters with colors redder and bluer than $(F475W-F814W)=1.5$.  

The choice of color separation above is an estimate based on the fact that 
the bright end of the color distribution of globular clusters around NGC 4874 can be 
fit with two Gaussians. These two Gaussians intersect at 
(F475W-F814W)=1.61 (Madrid et al.\ 2010). If one accounts for $\sim$0.1 mag for the blue tilt 
of bright, metal-poor globular clusters (e.g. Peng et al.\ 2009), the dividing line 
between red and blue globular clusters can be set at $(F475W-F814W)=1.5$. Given 
that the vast majority of globular clusters in our sample (98.3\%) have colors between 
$0.5<(F475W-F814W)<2.5$, the value $(F475W-F814W)=1.5$ is also a simple even separation.

The surface density distribution of red and blue globular clusters is presented 
in Figure \ref{redblue}. Red globular clusters 
are 52\% of the sample, the remaining 48\% are blue. In Coma, red globular clusters reach much higher 
surface densities than blue ones. Indeed, the highest surface density for blue clusters
is 0.10 GC/$\arcsec^2$ while this value reaches 0.41 GC/$\arcsec^2$ for red clusters.
Red GCs define high surface density clumps around the main subgroups of globular 
clusters associated with NGC\ 4889, NGC\ 4874, and IC\ 4051. 


\section{Summary and Final Remarks}

Globular clusters are distributed following three main overdensities in the core of Coma.
The most extended overdensity is associated with NGC\ 4874, while the highest surface density
of globular clusters is cospatial with NGC\ 4889. The second highest surface density 
of globular clusters coincides with IC\ 4051. The spatial distribution of globular clusters
around NGC\ 4874, NGC\ 4889, and IC\ 4051 is five to six times more extended than the 
corresponding galaxy light. 

The wide-field map of globular clusters presented here can be used in future work 
to investigate the evolutionary history of the Coma cluster by comparing it with 
numerical simulations like the one recently published by \citet{ramos2018}.

This work shows what can be achieved when combining high sensitivity, high resolution, and 
a wide field of view. WFIRST will have a sensitivity comparable to the HST and a much wider field
of view that will allow us to image the entirety of galaxy clusters, such as Coma, with 
a single pointing.


\acknowledgments

We would like to thank the anonymous referee for  prompt and constructive reports.
Based on observations made with the NASA/ESA Hubble Space Telescope, obtained at the Space 
Telescope Science Institute, which is operated by the Association of Universities for Research 
in Astronomy, Inc., under NASA contract NAS5-26555. These observations are associated with 
programs GO 10861, 11711, 12918. This research has made use of the NASA Astrophysics Data 
System Bibliographic services (ADS) and the NASA/IPAC Extragalactic Database (NED). AG acknowledges 
funding of the National Science Foundation through the Research Experience for Undergraduates (REU) 
program, grant AST-1062976. CO acknowledges the support of an Australian Gemini Undergraduate Summer 
Studentship from the Australian Astronomical Observatory funded by Astronomy Australia Ltd.

This work uses SDSS imaging data. Funding for the SDSS and SDSS-II has been provided by the Alfred P.\ Sloan 
Foundation, the Participating Institutions, the National Science Foundation, the U.S.\ Department of 
Energy, the National Aeronautics and Space Administration, the Japanese Monbukagakusho, the Max Planck 
Society, and the Higher Education Funding Council for England. The SDSS website is http://www.sdss.org/.


\appendix

This appendix gives additional details on how the final sample of globular clusters is built.
Table \ref{selectioncriteria} summarizes the main steps taken to create the final catalog of globular clusters in 
Coma. The selection of globular cluster candidates is based on their photometric properties
and sizes.

\section{Source finding}

The first step is to run \texttt{daofind}, within \texttt{daophot}, to detect all sources at 1.8$\sigma$ 
above the background. This value is determined through an iterative process. It is fine-tuned, by trial and error,
to over-detect sources and ensure that all potential globular clusters are selected. False positives are removed 
\textit{a posteriori}, see Fig.\ 7. The second step consisted of identifying and removing spurious detections along the edges, 
and along the ACS chip gap. 

For some bright galaxies \texttt{daofind} makes a large number of source detections around their 
centers. A steep gradient of galaxy light is often mistaken as a source by \texttt{daofind}.
To circumvent this \texttt{daofind} issue, all detections within a few arcseconds of the 
center of bright galaxies are removed. Detections of true globular cluster candidates
are then re-added by hand. For these inner regions of bright galaxies, we also use the median-subtracted 
images to better identify globular cluster candidates on the \texttt{ds9} display. The median-subtracted images are 
produced using the commands \texttt{median} and  \texttt{imarith}, the latter to subtract the median 
from the original image. We run \texttt{find} on the original and median-subtracted images, both sets of 
detections are then cross-matched to create a master list of detections.

\section{Color selection}

Once the photometry is obtained for the preliminary list of detections, a preliminary color-magnitude diagram 
is made. The CMDs of extragalactic globular cluster systems are well studied, beginning with the 
work done by Larsen et al. (2001).  On that earlier work, Larsen et al. (2001) established a selection 
criteria for the color of globular clusters of $0.7< (V-I)<1.45$ (translating into $-2.2 <[Fe/H] <0.2$). 
\citet{peng2006} selected objects with the color range $0.5< (F475W-F850LP)<2.0$ for their study of color 
distributions of globular clusters in Virgo.


\begin{center}
\begin{deluxetable}{clcc}
\tablecaption{Selection Criteria} 
\tablehead{
\colhead{Step} & \colhead{Selection Criterion} & \colhead{Visual} & \colhead{Number}\\
\colhead{} & \colhead{} & \colhead{Inspection?} & \colhead{of Sources}\\
}
\startdata
\cutinhead{{\it Detection}}
1  & First round of detections with DAOPHOT & Yes & 95785 \\
2  & Remove spurious detections around  edges and ACS chip gap & Yes & 80618 \\
3  & Remove false detections near galaxy centers (inner few arcseconds) & Yes & 62882 \\
\cutinhead{{\it Color selection}}
4  & Run  \texttt{daophot} to create a Color-Magnitude Diagram & No & -- \\
5  & Inspect color outliers (i.\ e. outside $0.7<(F475W-F814W)<2.0$) & Yes & 50669 \\
\cutinhead{{\it Size selection}}
7  & Run \texttt{source extractor} to determine sizes & No & -- \\
8  & Match sourcextractor and \texttt{daophot} catalogs, create size-magnitude plot & No & -- \\
9  & Second inspection of sources outside  $1.8>FWHM>3$ pixels & Yes & 30328 \\
\cutinhead{{\it Final verification}}
10 & Final visual verification & Yes & 23271 \\
\hline
\multicolumn{4}{c}{}\\ 
\multicolumn{4}{c}{Final sample           22426}\\   
\enddata
\label{selectioncriteria}
\end{deluxetable}
\end{center}


\begin{figure}
 \plotone{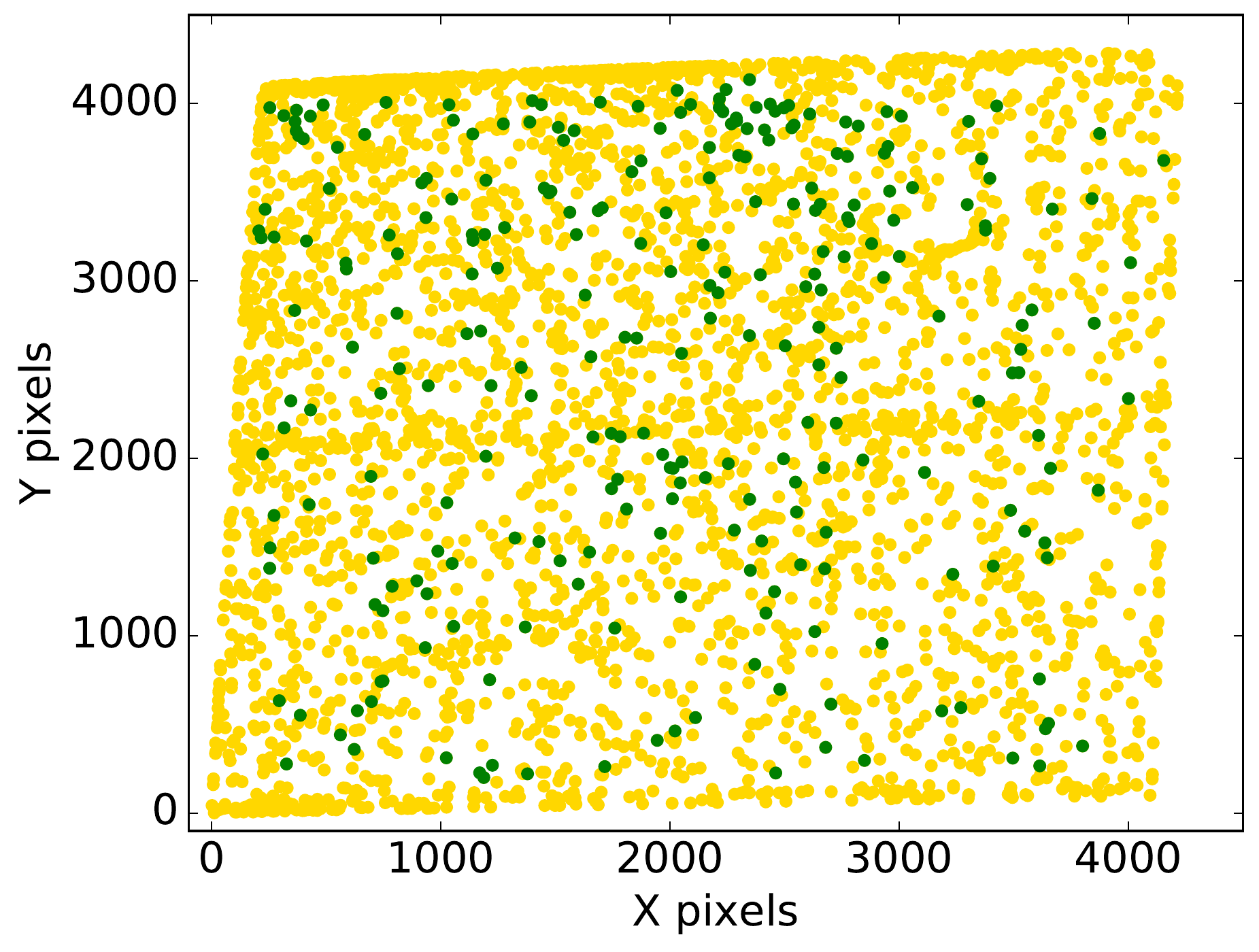}
 \caption{ACS pointing where we show all our original detections (in yellow) and the remaining 
 globular clusters (in green) after down selection.
 \label{detections}}
\end{figure}

An initial color-magnitude diagram allows us to build a list of objects that need secondary 
visual inspection due to their suspicious colors or magnitudes, such as being very red or blue, 
or overly bright or faint. Candidates with colors redder than $(F475W-F814W)>2.0$ and bluer
than  $0.7<(F475W-F814W)$ received a secondary inspection. Suspicious detections are displayed on the 
images and evaluated on individual grounds. To give a crude example, globular clusters in Coma should not 
have spiral arms, or show visible elongation on the ACS images. Some background galaxies can be easily picked 
up because of their colors. As shown in Figure \ref{cmd}, the vast majority of globular clusters have colors in the 
following range: $0.5<(F475W-F814W)<2.5$. Sources with colors redder than $(F475W-F814W)>3.0$ are also excluded
from the master catalog given that these are likely high-redshift objects.
Deriving flux measurements also allows us to remove a few detections 
that have bad (e.g.\ negative) fluxes. 

\bigskip
\begin{figure}
 \plotone{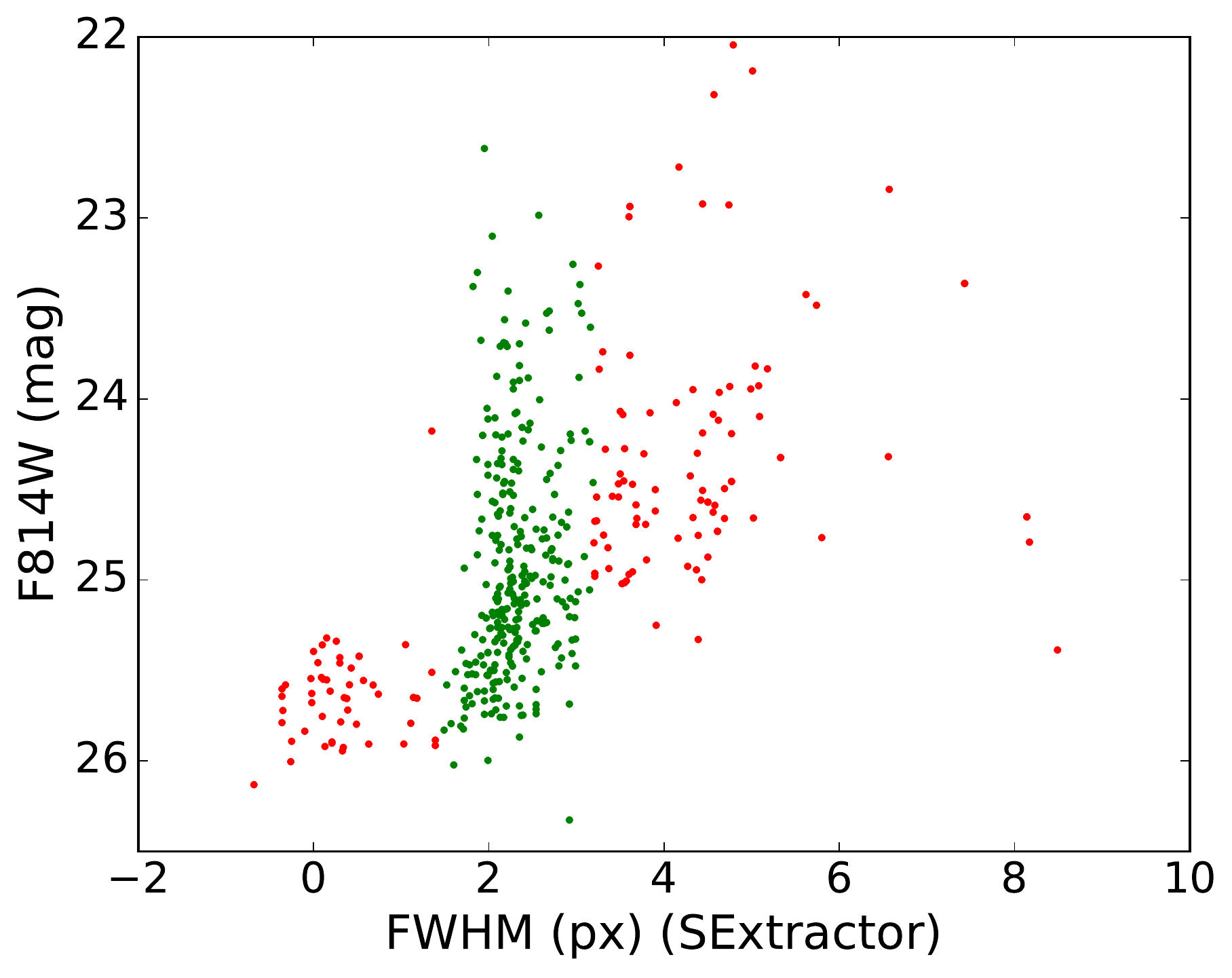}
 \caption{Size-magnitude diagnostic plot for a sample pointing. Red sources are displayed 
 on the images for further scrutiny. Sources with a FWHM$\sim$0 are almost invariably cosmic 
 rays and other detector blemishes. 
 \label{sizemagnitude}}
\end{figure}

\section{Size selection}

As mentioned above, morphology is a powerful tool to discriminate between globular clusters 
and interlopers, especially when using high-resolution data. Before we do our morphological analysis,
a PSF is derived by identifying several stars in the fields, obtaining their photometry 
and then running the \texttt{PyRAF} tasks \texttt{pstselect} and \texttt{psf}. The ACS/WFC
PSF has a FWHM of $\sim$2 pixels. At the distance of Coma, globular clusters appear
as point sources on the ACS images. Most globular clusters have radii of $\sim$3 pc, 
that is, are unresolved at the distance of Coma with a FHWM of $\sim$2 pixels. Great 
care is taken to preserve UCDs in our sample. Indeed, UCDs are slightly resolved at the 
distance of Coma with FWHM between $\sim$2 and $\sim$3 pixels.

To help with narrowing the list of globular cluster candidates, we derive the sizes of 
our detections using \texttt{Sextractor} \citep{bertin1996}. With the morphological
information, a list of sources that are either too small (FWHM~$<$1.8 px) or too big (FWHM$>$~3.0 px) 
is created for inspection on the images; see the size-magnitude diagram shown in Fig.~\ref{sizemagnitude}. 
As mentioned above, globular clusters have sizes similar to the instrumental PSF.
Most of the detections with sizes smaller than the PSF (e.\ g.\ $\sim$0.1 pixels) are cosmic rays.
Similarly, large objects are usually background galaxies that have radii of tens or hundreds of
pixels. \citet{madrid2010} gave an example of background galaxies, present in this dataset,
that can be easily spotted by running a basic size estimate.


\section{Final verification}

We then conduct a careful manual inspection of  every remaining GC candidate. This is done 
with the help of the \texttt{PyRAF} task \texttt{imexam}, which we use to display intensity 
contours and 3D mapping (surface wire-plot) of the point sources. We also use \texttt{imexam} 
to derive the radial profile of globular cluster candidates and obtain an approximate size, 
which is compared to the PSF. Figure \ref{detections} shows a sample ACS pointing where we show
all the original and final detections.

\end{document}